\documentclass[twocolumn,showpacs,amsmath,amssymb,pr,superscriptaddress]{revtex4-1}
\usepackage{graphicx}
\usepackage{ulem}
\usepackage{bm}
\usepackage{xcolor}
\usepackage[unicode=true,colorlinks=true,urlcolor=blue,citecolor=blue]{hyperref}

\begin{document}

\title{Electronic band structure in $n$-type GaAs/AlGaAs wide quantum wells in tilted magnetic field}

\author{I.~L.~Drichko}
\author{I.~Yu.~Smirnov}
\affiliation{Ioffe Institute, 194021 St. Petersburg, Russia}
\author{A.~V.~Suslov}
\affiliation{National High Magnetic Field Laboratory, Tallahassee, FL 32310, USA}
\author{M.~O.~Nestoklon}
\affiliation{Ioffe Institute, 194021 St. Petersburg, Russia}
\author{D.~Kamburov}
\author{K.~W.~Baldwin}
\author{L.~N.~Pfeiffer}
\author{K.~W.~West}
\affiliation{Department of Electrical Engineering, Princeton University, Princeton, NJ 08544, USA}
\author{L.~E.~Golub}
\affiliation{Ioffe Institute, 194021 St. Petersburg, Russia}

\begin{abstract}
Oscillations of the real component of AC conductivity $\sigma_1$ in a magnetic field were measured in the n-AlGaAs/GaAs structure with a wide (75 nm) quantum well by contactless acoustic methods at $T$=(20-500)~mK. {In a wide quantum well, the electronic band structure}
is associated with the two-subband electron spectrum, namely the symmetric (S) and antisymmetric (AS) subbands formed due to electrostatic repulsion of electrons. A change of the oscillations amplitude in tilted magnetic field observed in the experiments occurs due to crossings of Landau levels of different subbands (S and AS) at the Fermi level. The theory developed in this work shows that these crossings are caused by the difference in the cyclotron energies in the S and AS subbands induced by the in-plane magnetic field.
\end{abstract}

%\date{\today}

\maketitle

\section{Introduction} \label{introduction}

Transport investigations of two-dimensional (2D) high-mobility structures demonstrate a variety of effects taking place in different ranges of magnetic fields and temperatures. In moderate fields, the Shubnikov-de Haas oscillations are observed, while in stronger fields, the integer and fractional quantum Hall effects occur~\cite{Shayegan_review_2006}. In some systems the Wigner crystal is formed~\cite{Shayegan_review_2006}. The most interesting results are mainly obtained on $n$-type GaAs/AlGaAs structures due to high mobility of carriers of up to $\sim 10^7$~cm$^2$/(Vs).

Since the general description of the phenomena observed in 2D systems in high magnetic fields is well developed, the interest is now shifting to more exotic systems and situations. At large half-integer filling factors $\nu >5/2$ and low temperatures $T<150$~mK the
formation of the stripe phases due to strong electron-electron interaction has been predicted~\cite{stripe_prediction1,stripe_prediction2,stripe_prediction3}. This phase is characterized by a strong anisotropy of longitudinal conductivity $\sigma_{xx}$, and the stripe formation explains the oscillations in ultra high-mobility heterojunctions~\cite{exp_stripe}. Recently it has been demonstrated experimentally that the stripes can be re-oriented by the  in-plane component of a magnetic field, i.e., in a tilted magnetic field~\cite{stripes_parallel_field}.

The conductivity in tilted magnetic fields is rather non-trivial in 2D structures with a wide quantum well (WQW). In such systems electrons are localized near the interfaces due to electrostatic repulsion. As a result, a formed two-layer system is similar to a  pair of quantum wells. Due to an inter-layer tunneling the electronic energy structure consists of two subbands, the symmetric (S) and antisymmetric (AS), separated by an energy gap $\Delta_\text{SAS}$~\cite{shayegan_review}.
In low magnetic fields conductivity oscillations caused by elastic scattering between the S and AS subbands were observed which are different from the Shubnikov-de~Haas oscillations~\cite{Drichko_2018}.
The presence of two conductivity channels in WQWs results in conductivity magnetooscillations corresponding to filling factors $\nu$ with even denominators.
At small filling factors $\nu =1/2, 3/2$ the fractional quantum Hall effect has been observed in some WQWs. A phase diagram has been constructed demonstrating conditions of stable observation of the fractional quantum Hall effect at $\nu =1/2$ at various electron densities, quantum well widths, and values of $\Delta_\text{SAS}$, see Ref.~\cite{FQHE_diagram} and references therein. Also a gas of composite fermions has been observed in WQWs and investigated in detail, see, e.g., Ref.~\cite{comp_ferm}.

The conductivity magnetooscillations can be strongly affected by the in-plane  field component.
For example, in systems with one 2D subband, the Landau level crossing in tilted magnetic fields occurs because the orbital splitting is determined by the perpendicular field component, while the spin splitting is given by the total magnetic field. It results in crossing of opposite-spin sublevels pertaining to different Landau levels. This crossing allows determining of the $g$-factor from the magnetoconductivity measurements by a method proposed long time ago~\cite{g-factor} and widely used up to nowadays.
The situation is much  richer in systems with two electronic subbands, in particular in WQWs, due to a presence of the $\Delta_\text{SAS}$ gap.
It has been pointed out in Refs.~\cite{Shayegan_2010,Shayegan_2015} that the crossing of Landau levels from different subbands (S and AS) can occur in this case in a tilted field, since  the value of $\Delta_\text{SAS}$ can depend on the in-plane  field component.
Calculations of the energy spectrum in parallel field $\textbf B_{\parallel}$ show that the gap $\Delta_\text{SAS}$ increases at small $B_{\parallel}$ while in stronger fields $B_{\parallel} >2$~T, the inter-layer coupling is suppressed, and $\Delta_\text{SAS} \to 0$~\cite{orbits_delta}.
This result demonstrates a great potential of experiments on high-mobility WQWs in tilted magnetic fields.

There are interesting studies of conductivity in bilayer systems in tilted magnetic fields. Such works were made on double quantum well systems (Refs.~\cite{new12,new13,new14,new15,new16,new17}) as well as on a WQW~\cite{orbits_delta,new18,MISO_tilted_field_2,Bykov2019}. The magnetoresistance studies in most of these papers were carried out in moderate magnetic fields where SdH and intersubband oscillations coexist and produce a beating pattern. By contrast, in the system under present study $\Delta_\text{SAS}= 0.42$~meV is very small and therefore intersubband scattering affects the conductivity only in small magnetic fields $B <  0.3$~T~\cite{Drichko_2018}. In this study we analyze results obtained in higher fields 0.5~T$<B_{\perp} <  1.5$~T, i.e., in the range where no beatings were observed.

We show that in a tilted field minima of these oscillations are converted to maxima due  to  $B_{{\parallel}}$ induced crossings of the Landau levels caused by the complicated energy scheme of WQWs.
We developed a theory demonstrating that the crossings are caused by the effect of the parallel field on the cyclotron energies in S and AS subbands.

\section{Experiment} \label{experiment}

We studied a multilayered  $n$-GaAlAs/GaAs/GaAlAs structure with a 75~nm wide GaAs quantum well.
The WQW was $\delta$-doped on both sides and located at the depth  $\approx 197$~nm below the surface of the sample. By illuminating the sample with infrared light of an emitting diode when cooling the heterostructure down to 15 K we achieved the electron mobility $\mu=2.2\times 10^7$~cm$^2$/(Vs) and density ${n_e=3\times 10^{11}}$~cm$^{-2}$ (at $T=0.3$~K). In the absence of a magnetic field the S-AS energy splitting in this structure was $\Delta_\text{SAS}=0.42$~meV~\cite{Drichko_2018}.

In our contactless acoustic experiments a surface acoustic wave (SAW) propagates along a surface of a piezoelectric crystal LiNbO$_3$ while the sample is slightly pressed onto this surface on which interdigital transducers are deposited to generate and detect the wave. An ac electric field accompanying the SAW penetrates into the 2D channel located in the semiconductor structure. The field produces electrical currents which, in turn, cause Joule losses. As a result of the interaction of the SAW electric field with charge carriers in the quantum well the SAW attenuation $\Gamma$ and its velocity shift $\Delta v/v$ are governed by the complex high-frequency conductance, $\sigma^{AC}=\sigma_1(\omega)-i\sigma_2(\omega)$. The SAW technique has shown to be a very efficient tool as it allows studying both real $\sigma_1$ and imaginary $\sigma_2$ components of the high-frequency conductance without any needs for electrical contacts~\cite{Drichko2000}.

\begin{figure}[h]
\centering
\includegraphics[width=\linewidth]{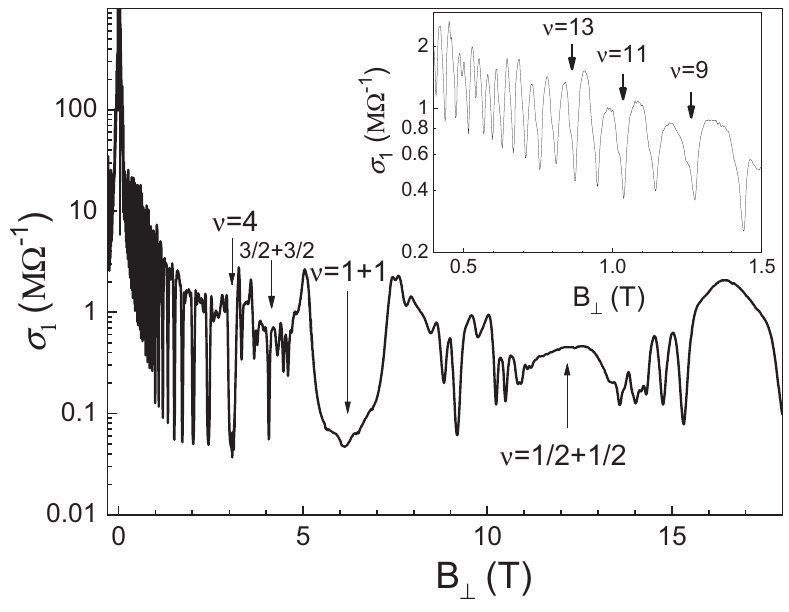}
\caption{Dependence of the real part of the ac conductivity $\sigma_1$ on magnetic field $B_\perp$ at $T=20$~mK. SAW frequency $f=30$~MHz.
%The filling factors are denoted near oscillations.
Inset: $\sigma_1$ in the range $B_\perp=0.5\ldots 1.5$~T at $T=310$~mK.
}
\label{fig1}
\end{figure}

The measurements in a magnetic field perpendicular to the sample plane were carried out in a dilution refrigerator. Dependences of the absorption coefficient $\Gamma$ and  the velocity change $\Delta v/v$ on the magnetic field at $B\leq 18$~T were measured in the temperature interval 20-500~mK. The SAW frequency was changed from 28 to 307~MHz. The ac conductivity $\sigma^{AC}$ was deduced with help of the expressions published in  Ref.~\cite{Drichko2000}. The measurements in tilted magnetic fields were done at SAW frequency of 30~MHz, in the same magnetic field range of 18~T, and at the temperature 0.3~K in a $^3$He cryostat. Samples were mounted on a one-axis rotator, which enabled us to change the angle between the normal to the structure and the magnetic field.
In the present work, we analyzed dependences of the real component of the conductivity $\sigma_1 \gg \sigma_2$ (in this case $\sigma_1\cong\sigma^{DC}$) on the magnetic field.
Figure~\ref{fig1} shows the dependence of the real component of the conductivity $\sigma_1$ on the perpendicular magnetic field at $T=20$~mK. In high magnetic fields conductivity oscillations corresponding to the fractional quantum Hall effect are observed at $n_e=3 \times 10 ^{11}$ cm$^{-2}$.
%
%\sout{In our structures total concentration was $n=3\times 10^{11}$~cm$^{-2}$, while the difference $\Delta n=1.2\times 10^{10}$~cm$^{-2}$.}
In this article we report on our studies of  oscillations in the range $B_\perp=0.5\ldots 1.5$~T. Dependence $\sigma_1(B)$ in this field domain at $T=310$~mK is illustrated in the inset to Fig.~\ref{fig1}.

In the double-layer structure formed in WQWs the total electron density is {\it usually} determined from the position of the minimum (maximum in the case of $\nu$=1/2) of the FQHE resistance (conductivity) oscillations in strong magnetic fields, and the difference of the charge density in the layers is derived using Fourier analysis of magnetoresistance in fields $B < 0.4-0.5$~T. However, in the case of our multilayer sample Fig.~\ref{fig1} requires a special discussion. The oscillations pattern near $B=$12.35~T looks the same as usual for $\nu$=1/2, which corresponds to concentration $n_e \sim 1.5 \times 10^{11} $~cm$^{-2}$. At the same time analysis of oscillations in the field $B<$3~T gives $n_e \sim 2.7 \times 10^{11} $~cm$^{-2}$. We believe that in large magnetic fields, the layers in our wide quantum well are practically independent (due to small $\Delta_\text{SAS}$=0.42 meV), and the conductivities from different layers with almost the same concentration are simply added. The appearance of oscillations at $B=$12.35~T corresponds to $\nu$ = 1/2 for each layer with a concentration $\sim 1.5 \times 10^{11} $~cm$^{-2}$.
One needs to draw the attention that determination of the electron density from the oscillation position in a magnetic field gives different values for low and high fields~\cite{densityfield}. In the present work at $B>$4~T $n_e=3 \times 10 ^{11}$ cm$^{-2}$, while in the region under this study $B<$2~T $n_e=2.7 \times 10 ^{11}$ cm$^{-2}$. The latter  value of the total concentration of electrons  is yielded also from both Fourier analysis and the dependence of $\rho_ {xy}$ at classically strong fields $B<$1~T.

\begin{figure}[h]
\centering
\includegraphics[width=\linewidth]{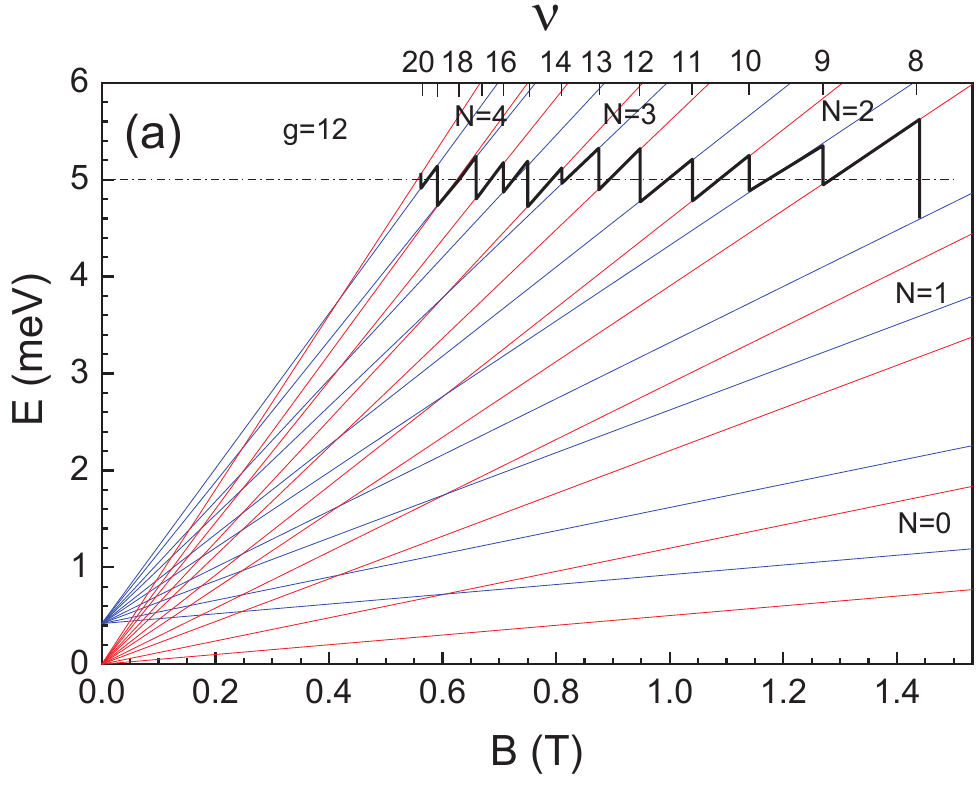}
\\
\includegraphics[width=\linewidth]{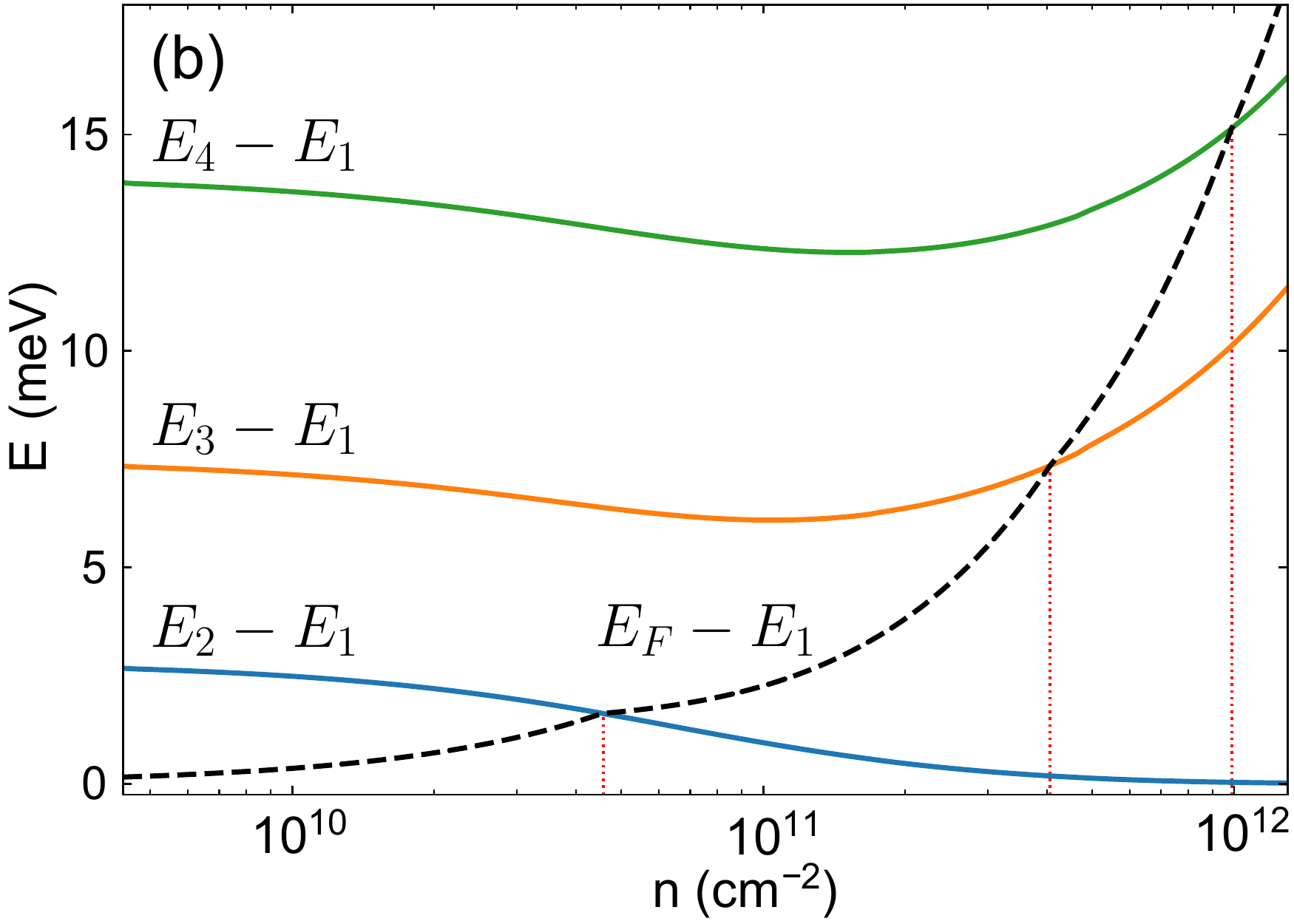}
\caption{
(a) Dependences of the energies of the spin split ($g$=12) Landau levels in the WQW on magnetic field $B$ (bottom) and filling factors $\nu$ (top) with the tunneling splitting $\Delta_\text{SAS}=0.42$~meV at $\textbf B_{\parallel}$=0. The Landau levels of the lower (S) and higher (AS) subbands are shown by red and blue colors, respectively. Black dash-dot line is the Fermi level at $B=0$ corresponding to the total electron density $n_e=2.7\times 10^{11}$~cm$^{-2}$. The black solid line is the Fermi energy dependence on $B$. (b) Dependencies of the size-quantization levels from $E_2$ to $E_4$ and the Fermi level $E_F$ position on the electron density in the 75~nm WQW. All energies are counted from the ground level $E_1$. Vertical dotted lines indicate the values of density where filling of the next size-quantized levels starts.
}
\label{fig2}
\end{figure}

In order to understand the nature of the oscillations we consider a fan of Landau levels for the two-subband system under study. Figure~\ref{fig2}(a) represents the Landau level fan diagram for our two subbands with spin splitting ($g$=12). Indices N in Fig.~\ref{fig2}(a) represent Landau levels counted for each subband independently, while filling factor $\nu$ is equal to total amount of Landau levels located below the Fermi level. Symmetric (red lines) and antisymmetric (blue lines) are the Landau levels with the gap $\Delta_\text{SAS}=0.42$~meV reported for this sample in Ref.~\cite{Drichko_2018}. The Fermi energy $E_{F}=5$~meV (at $B$=0) corresponds to the total concentration $n_e=2.7\times 10^{11}$~cm$^{-2}$ in the two-subband balanced system. The positions of size-quantized levels and the Fermi level in the 75~nm-wide quantum well are presented in Fig.~\ref{fig2}(b). The energies are found by solving the coupled Schr{\"o}dinger and Poisson equations at $T$=0 in the effective mass method with the following parameters: the barrier height 370~meV and the electron effective mass $m=0.067 m_0$. It is clear from Fig.~\ref{fig2}(b) that two levels of size quantization are occupied at $5 \times 10^{10} < n_e < 4 \times 10^{11}$~cm$^{-2}$.

Our analysis (not presented here) demonstrates that in the minima shown in Fig.~\ref{fig1} conductivity has an activation character. We compared the positions of the conductivity minima with the Fig.~\ref{fig2} fan diagram and found that these minima are related to the charge carrier transitions between different spin-split subbands S and AS. Therefore, the activation energies for different $\nu$ are determined by steps in the Fermi energy dependence on magnetic field. In Fig.~\ref{fig2}(a) vertical lines in the Fermi energy dependence represent these steps and are drawn at the magnetic field values, where oscillation minima were observed. It follows from Fig.~\ref{fig2}(a) that there are four types of oscillation minima at $B_{{\parallel}}$=0 associated with transitions between subbands. One of them are determined by the activation energy equal to the difference $E_{N+1,\uparrow}^{S}-E_{N,\downarrow}^{AS}$ {($\nu$= 8, 12, 16, 20)} and the rest demonstrates other possible transitions between AS and S related levels: $E_{N,\downarrow}^{S}-E_{N,\uparrow}^{AS}$ ($\nu$=10, 14, 18), $E_{N,\downarrow}^{AS}-E_{N,\downarrow}^{S}$ ($\nu$=11, 15), and $E_{N,\uparrow}^{AS}-E_{N,\uparrow}^{S}$ ($\nu$=9, 13).

The oscillation pattern in the WQW is even more intriguing in tilted magnetic fields. The results of corresponding measurements are presented in Fig.~\ref{fig4}. This figure shows that at some tilt angles of the magnetic field the minima of the conductivity oscillations are changed to maxima. The experimental results presented in Fig.~\ref{fig4} were obtained at the in-plane component of the magnetic field $\textbf B_{\parallel}$ oriented along the SAW propagation direction. We also performed experiments in another configuration and observed that the minima of conductivity oscillations transformed to maxima at the same angles when $\textbf B_{\parallel}$ was oriented perpendicular to the SAW wave vector.

{As mentioned above, the temperature dependence of the conductivity in minima has the activation character. Thus, the conductivity $\sigma \propto \exp{[-\Delta_a/(2k_{B}T)]}$, and the conductivity minima transforms into maxima when $\Delta_a \to 0$, i.e., at crossings of Landau levels. The nature of these crossings in the two-subband system under study is discussed in the next section.

\begin{figure}[h]
\centering
\includegraphics[width=\linewidth]{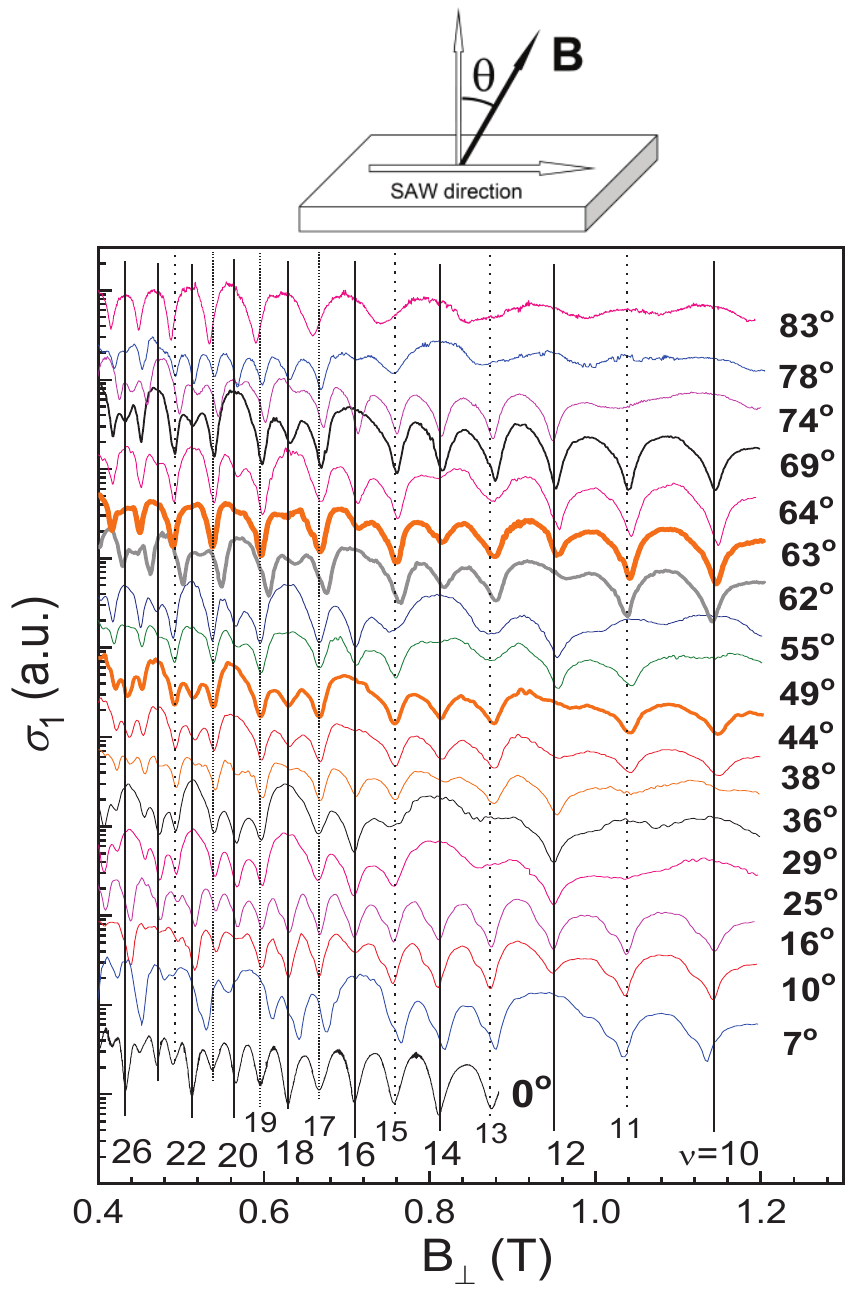}
\caption{Dependence of the conductivity $\sigma_1$ on the normal component $B_\perp$ at different angles of the magnetic field tilt  at $T=310$~mK, $f=30$~MHz. The numbers below the minima  denote
 the numbers of filled Landau levels under the Fermi level, $\nu$.
Traces are offset for clarity.
}
\label{fig4}
\end{figure}

\section{Theory}

There is a big theoretical activity devoted to study of bilayer systems including WQWs in the presence of tilted magnetic field~\cite{new14,new17,new18,Arapov}.
Most of these works are aimed at numerical calculation of Landau levels and their change by the in-plane field component.
%are performed on a more complicated theoretical level with emphasis on numerical calculation of Landau levels.
By contrast, we perform semiclassical analysis and derive analytical expressions for the change of the electron effective mass by the in-plane magnetic field component. We show below that these expressions are appropriate for description of our experimental data. Besides, due to weak coupling of the layers in the studied structure, the effect of the in-plane field on the S-AS tunneling splitting important in previous studies does not play a big role in our analysis. Instead,
%: the results of numerical calculations presented in the inset to Fig. 5(b) show that even enhanced in the tilted magnetic field S-AS splitting is still small, and
the crossing of the Landau levels is due to a change of the energy dispersion rather than due to modification of tunneling.

The Hamiltonian of a doped WQW in a parallel magnetic field is given by
\begin{equation}
\label{H}
 {\cal H} =  -{\hbar^2 \over 2 m}{d^2 \over dz^2} + {1 \over 2 m} \left(\textbf p - {e\over c} \textbf A \right)^2 +e\phi(z) .
\end{equation}
Here $z$ is the axis normal to the WQW, $\textbf p$ is the in-plane electron momentum, $\textbf A=z(B_y,-B_x,0)$ is the vector potential of the magnetic field, and $\phi(z)$ is the electrostatic potential.

We performed self-consistent calculations of the electrostatic potential and electron wavefunctions. First, the wave functions are calculated in the tight-binding approach~\cite{tight-binding}.
Then, the electron wave functions are used to calculate the electron density distribution
in the quantum well as explained in Refs.~\cite{Drichko_2018,Schurova_Galperin}.
The value of the total electron density extracted from our experiment is
${n_e=2.7\times10^{11}}$cm$^{-2}$.
The electrostatic potential $\phi(z)$
corresponding to the charged QW is found from the numerical solution of
the Poisson equation at zero in-plane momentum $\textbf p=0$.
{We account for the in-plane magnetic field $B_{\parallel}$ through the effective potential
$\phi_{B} = e(B_{\parallel} z/c)^2/2 m $, see Eq.~\ref{H}. We add the term $e\phi(z)+e\phi_{B}$}
to the structure potential and compute the next approximation for the electron wave functions of the levels in the WQW.
The procedure is repeated until the self-consistency of the electron wave functions and
electrostatic potential is reached.
In accordance with previous studies~\cite{shayegan_review,Drichko_2018} we obtained that
the carriers are located near the WQW edges in the narrow layers separated by a distance $d=56$~nm.
At $B_{\parallel}=0$ $\Delta_\text{SAS}=0.42$~meV, and the in-plane magnetic field changes this value due to the diamagnetic term $e\phi_{B}$. We discuss this dependence in the next section.
Here we note that $\Delta_\text{SAS}$ has weak dependence on $B_{\parallel}$ increasing by just about 40$\%$ at $B_{\parallel}$=2~T.

\begin{figure}[h]
\centering
\includegraphics[width=\linewidth]{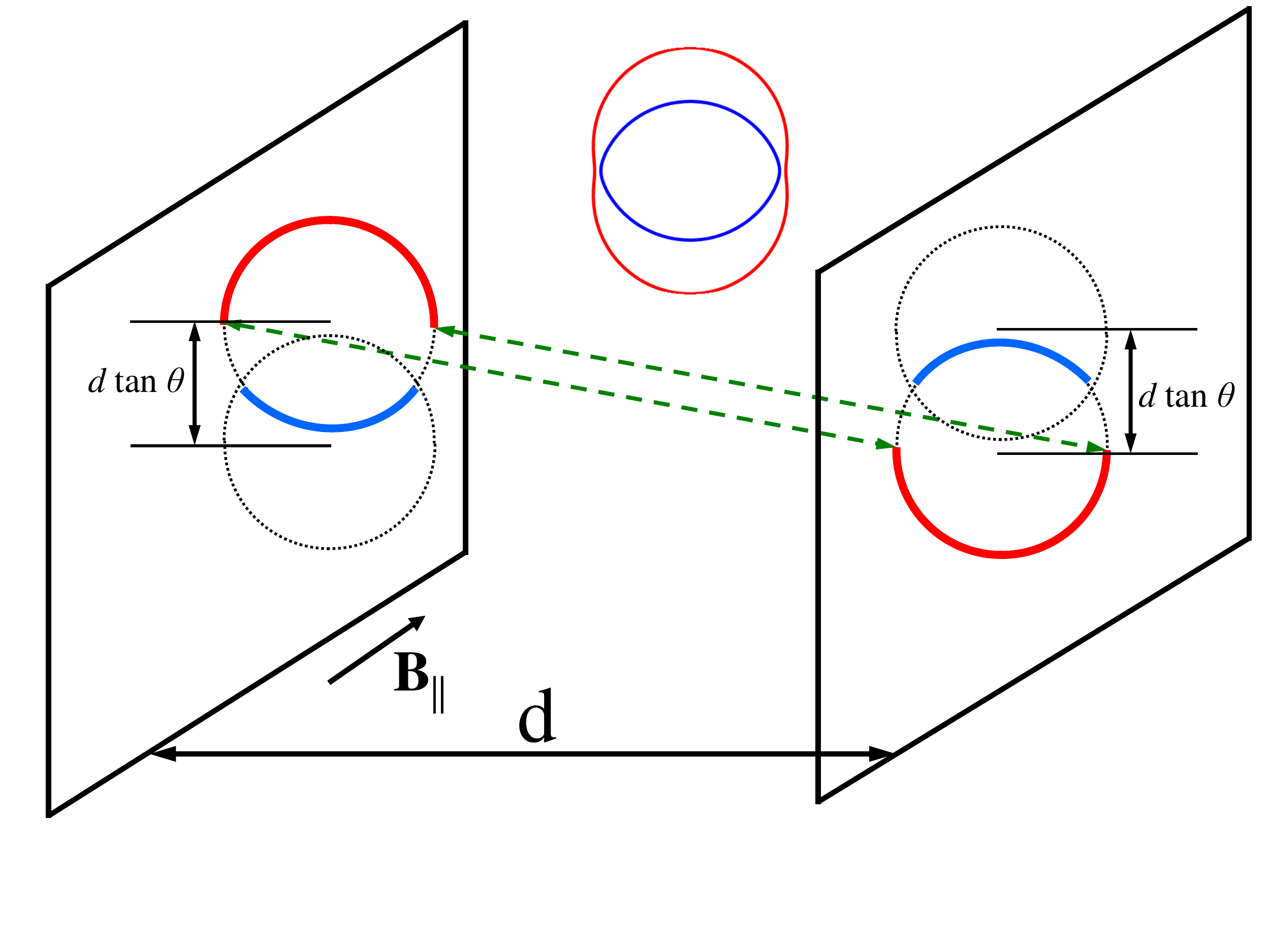}
\caption{Scheme of cyclotron trajectories in WQW representing a two-layer system with the interlayer separation d. Green dashed lines show the tunneling processes. Due to the separation of the cyclotron orbit centers in the layers, the outer (red) trajectory has a larger cyclotron period than the inner (blue) trajectory.
}
\label{Orbits}
\end{figure}

Not taking into account $\Delta_\text{SAS}$, we consider the states in the left (L) and right (R) layers separately
assuming the layers' width to be infinitely small. Then, in the presence of the in-plane magnetic field $\textbf B_{\parallel}$,
the energy dispersions of the L and R states
are anisotropic:
\begin{equation}
\label{en_spectrum}
E_\text{L,R}(\bm p)
={p^2\over 2m} +  {1\over 2m} \left({eB_\parallel d\over 2c} \right)^2 \pm {e  d\over 2mc} (\bm B_\parallel \times \bm p)_z,
\end{equation}
Here the second term is the diamagnetic shift equal for both states, and the separation of the subbands' minima in the momentum space, $\Delta p = eB_{\parallel} d/c$,
is given by the last term.
In a tilted magnetic field this separation results in the difference of centers of cyclotron circles $\Delta r = d \: B_{\parallel} /B_\perp$, Fig.~\ref{Orbits}.

Now we take into account that,
due to tunneling, the carrier trajectories are not just two shifted circles but are more complicated ones.
This fact results in a difference of the cyclotron energies.
Indeed, the cyclotron period $T$ in the perpendicular field $B_\perp$ is related to the area in the momentum space bounded by the trajectory, ${\cal A}$, by
\begin{equation}
\label{period}
T = {c\over |eB_\perp|} {\partial {\cal A} \over \partial E}, \qquad
{\cal A}=\oint\limits d\varphi {p^2(\varphi) \over 2},
\end{equation}
where $E$ is the carrier energy and $\varphi$ is an angle between $\textbf p$ and $\textbf B_{\parallel}$.
The electron momentum absolute values at a fixed energy $E$ for $0<\varphi\leq \pi/2$ are found from Eq.~\ref{en_spectrum}
\begin{equation}
p_{L,R}(\varphi,E)=
\left| {eB_{\parallel} d\over 2c} \sin{\varphi}
\pm
\sqrt{2mE - \left({eB_{\parallel} d\over 2c} \right)^2 \cos^2{\varphi} } \right|,
\end{equation}
and the momenta at other values of $\varphi$ are obtained from this expression by appropriate rotations.
With account for tunneling the S and AS states are formed, and the electron trajectories are different from simple circles~\cite{new13,orbits_delta,MISO_tilted_field_2}.
%As a result, a half of a period the momenta are equal to $p_\{L}$, and the other half of period -- to $p_\{R}$.
The cyclotron orbits are shown schematically in Fig.~\ref{Orbits}. With account for tunneling the electrons spend by a half of their cyclotron periods in the left and right layers.
As a result, at $|\varphi|<\pi/2$ $p=p_{L}(\varphi)$ in the S subband and $p=p_{R}(\varphi)$ in the AS subband.
Substitution to Eq.~\ref{period} yields
\begin{equation}
{\cal A}_{S,AS} = 2\int_0^{\pi/2} d\varphi p_{L,R}^2(\varphi),
\end{equation}
and for the cyclotron periods in the S and AS subbands:
\begin{equation}
T_{S,AS}
={2\pi\over \omega_{c,0}}\left(1 \pm b\right), \quad b ={2\over \pi}\arcsin{\left({eB_{\parallel} d\over 2c\sqrt{ 2mE_{F}}} \right)}.
\end{equation}
Here $\omega_{c,0}=|eB_\perp|/mc$ is the cyclotron frequency in the absence of parallel field, and we took the electron energy $E$ equal to the Fermi energy $E_{F}$ assuming  $E_{F} \gg \Delta_\text{SAS}$.
As a result, we have different cyclotron frequencies
in two subbands:
\begin{equation}
\label{omegas}
 \omega_c^{S,AS} = {\omega_{c,0} \over 1\pm b}
%, \quad \hbar \omega_c^{AS}-\hbar \omega_c^{S} = {\hbar eB_\perp\over mc}{2b \over 1-b^2}
.
\end{equation}
The cyclotron periods are different due to tunneling and separation of the cyclotron trajectory centers.
% by the momentum $eB_{\parallel} d/c$.

We see that the cyclotron frequencies in the S and AS subbands depend on the tilt angle $\theta$ via $B_{\parallel}=B_\perp\tan{\theta}$.

\section{Discussion} \label{disc}

Figure~\ref{fig5} shows the angular dependencies of the real part of the conductivity on the tilt angle $\theta$ extracted from Fig.~\ref{fig4} for various filling factors $\nu$.
The dependence
$\sigma_1(\theta)$ at $\nu$=12, 16, 20 shown in Fig.~\ref{fig5}~(a)
%{\sout{ for the activation energy  equal to the difference  $E_{4,\uparrow}^{AS}-E_{5,\downarrow}^{S}$ (at $\theta=0$).}}
demonstrates that the oscillation minima change by maxima
at $\theta \approx 7^\circ,  44^\circ$ and~$60^\circ$.
Figure~\ref{fig5}~(c) shows that in the other series of oscillations at $\nu$=6, 10, 14, 18
%{\sout{ related with the Landau levels $N$S and $N$AS (at $\theta=0$),}}
the conductivity maxima take place at the angles $\theta \approx 34^\circ$ and $\theta \approx 54^\circ$.

The maxima in the conductivity in the activation regime are caused by intersection of some Landau levels at the Fermi level. These levels correspond
to S and AS subbands
and to opposite spin orientations.
Energies of the $N$th level of the S subband with spin down (${\downarrow}$) and $N'$th level of the AS subband with spin up (${\uparrow}$) are given by, [see Eq.~\ref{omegas}]:
\begin{align}
\label{E_n_n'}
&E_{N,\downarrow}^\text{S} = {\hbar \omega_{c,0}(N+1/2)\over 1+b} - {\Delta_\text{SAS}\over 2}- {g\mu_\text{B}B\over 2} , \nonumber
\\
&E_{N',\uparrow}^\text{AS} = {\hbar \omega_{c,0}(N'+1/2)\over 1-b} + {\Delta_\text{SAS}\over 2} + {g\mu_\text{B}B\over 2} .
\end{align}
Here $g$ is the Land\'e factor,  $\mu_{B}$ is the Bohr magneton, and $B$ is the total magnetic field.

Equation~\ref{E_n_n'} demonstrates three reasons for reconstruction of the energy levels by parallel magnetic field. First, the tunneling gap $\Delta_\text{SAS}$ changes in the parallel field~\cite{orbits_delta}. We have calculated the dependence of $\Delta_\text{SAS}$ on $B_{\parallel}$ as described in the previous section of this paper.
The results of calculations are shown in the inset to Fig.~\ref{fig5}~(d). We see that the dependence of $\Delta_\text{SAS}$ on the parallel field $B_{\parallel}$ is weak, so that $\Delta_\text{SAS}$ in the WQW under study is small in the whole range $0< B_{\parallel} <2$~T.
Our analysis shows that account for the dependence of $\Delta_\text{SAS}$ on the parallel field $B_{\parallel}$
shifts the crossing points of the Landau level for $\nu=20$ just by 0.2~meV which is 0.45~\% only.

Then, the Zeeman terms in Eqs.~\ref{E_n_n'} result in the dependence of the Landau levels on the tilt angle at a fixed perpendicular field via $B=B_\perp/\cos{\theta}$. However, for explanation of the level crossings at experimentally found angles one should take the values of $g \sim 30\ldots 40$ which are unrealistic.

Finally, we examined the dependences of the cyclotron energies in the S and AS subbands on $B_{\parallel}$.
According to Eq.~\ref{omegas} the cyclotron energy
%decreases in the S subband and increases in the AS subband with $\theta$.
 decreases/increases with $\theta$ in the S/AS subband.
It results in a dependence of the Landau level positions on $\theta$ via the factors $b(\theta)$, see Eqs.~\ref{E_n_n'}.
In Figs.~\ref{fig5}~(b),~(d) we plot the Landau level positions at the filling factor $\nu=16$ and $\nu=18$, respectively,  taking the inter-layer separation in the WQW under study as $d=56$~nm~\cite{shayegan_review,Drichko_2018} and   the Fermi energy in each subband as $E_{F}=5$~meV.
%Figure~\ref{fig5}~(b) demonstrates that
%at $B_\perp=0.565$~T, the
%%4th Landau level of the AS subband ($N'=4$)
%Landau levels 4AS and 5S
%intersect at the Fermi level
%%with the 5th Landau level of the S subband ($N=5$)
%at $\theta \approx 7^\circ$. At $\theta \approx 44^\circ$, the Landau levels
%%with $N'=3$ and $N=6$
%3AS and 6S cross the Fermi level.
%
%{\sout{We  plot the Landau levels as in Fig.~\ref{fig7} for
%various filling factors.
%From the analysis of odd $\nu=3,5,7,9$ and half-integer $\nu=11/2,13/2$
%we obtain
%that the Landau-level crossings occur at $\theta_c=30^\circ$ and~$54^\circ$, while for  even $\nu=6, 8, 10$ we get $\theta_c=7^\circ$, ~$44^\circ$, and~$59^\circ$.
%}}
%
The values of $\theta$ corresponding to conductivity maxima are shown in Fig.~\ref{fig5} by vertical lines.
We do not analyze the crossings at higher angles where $B_{\parallel} >1.5$~T because the developed theory does not describe this regime. The curves are plotted with account of the Zeeman splitting with $g <$0.

\begin{figure*}[t]
\centering
\includegraphics[width=\linewidth]{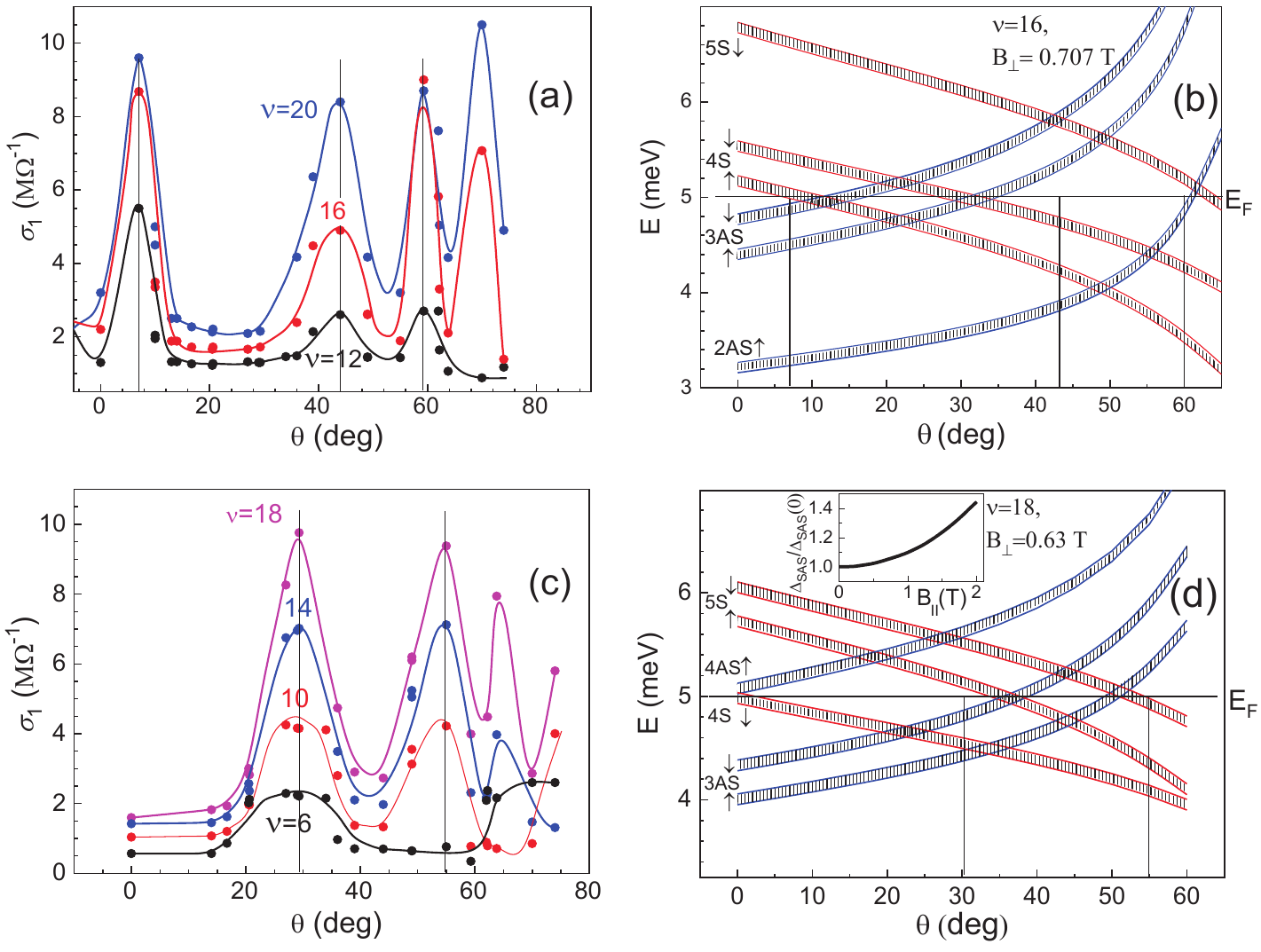}
\caption{
(a),(c): The conductivity as a function of the magnetic field tilt angle measured at different $\nu$, $f=30$~MHz.
$\nu$ are the numbers of filled Landau levels under the Fermi level.
(b),(d): Theoretical dependencies of energy of the spin split Landau levels  on the magnetic field tilt angle. The levels are calculated using Eqs.~\ref{E_n_n'} with account of the broadening via Eqs.~\ref{eqBroad} for subbands S (red) and AS (blue). The numbers before S(AS) denote LLs numbers.  The black solid horizontal line indicates the Fermi level $E_{F}$ at $B$=0. {Arrows show the spin orientation. The vertical lines illustrate experimental positions of the LLs crossings. Inset: The calculated dependence of $\Delta_\text{SAS}$ on $B_{\parallel}$.
}
\label{fig5}}
\end{figure*}

Two series of the observed oscillations
have maxima at different tilt angles.
At $\theta=0$, the oscillations presented in Fig.~\ref{fig5}~(a) are related to the activation gap
${\Delta_a=E_{5,\uparrow}^{S}-E_{4,\downarrow}^{AS} }$ ($\nu$=20),
${\Delta_a=E_{4,\uparrow}^{S}-E_{3,\downarrow}^{AS} }$ ($\nu$=16),
${\Delta_a=E_{3,\uparrow}^{S}-E_{2,\downarrow}^{AS} }$ ($\nu$=12).
Figure~\ref{fig5}~(b)  shows that { ($\nu$=16)  broadened levels 3AS$\downarrow$ and 4S$\uparrow$} are getting closer with tilting of the field, and at $\theta \approx 10^\circ$ they cross at the Fermi level resulting in
closing of the activation gap
and in rising of the conductivity from minimum to maximum.
With further increase of the tilt angle, the energies of these levels diverge,
and $\Delta_a$ increases until $\theta \approx 22^\circ$. At larger angles, the activation gap is formed from the converging energy levels 4S$\downarrow$ and 3AS$\uparrow$. In turn, these levels cross at $\theta \approx 31^\circ$, resulting in the next conductivity maximum. After the crossing they diverge as well, and the next conductivity maximum occurs at $\theta \approx 62^\circ$ when 5S$\downarrow$ crosses 2AS$\uparrow$. Analogous analysis performed  for $\nu$=20 and 12 demonstrate the same crossing angles.

The other series of oscillations presented in Fig.~\ref{fig5}~(c) has the conductivity maxima at the angles $\theta \approx 34^\circ$ and $\theta \approx 54^\circ$.
At $\theta=0$ these oscillations are related to
%the Landau levels $N$S and $N$AS.
the activation gap ${\Delta_a=E_{4,\uparrow}^{AS}-E_{4,\downarrow}^{S} }$ ($\nu$=18),
${\Delta_a=E_{3,\uparrow}^{AS}-E_{3,\downarrow}^{S} }$ ($\nu$=14),
${\Delta_a=E_{2,\uparrow}^{AS}-E_{2,\downarrow}^{S} }$ ($\nu$=10), and
${\Delta_a=E_{1,\uparrow}^{AS}-E_{1,\downarrow}^{S} }$ ($\nu$=6).
However,
%This is explained by
Fig.~\ref{fig5}~(d) , built for $\nu$=18,
%plotted by Eq.~\ref{E_n_n'}
%which
demonstrates
that the broadened levels 4AS$\uparrow$ and 4S$\downarrow$ diverge with increase of the tilt angle. At $\theta >15^\circ$ the activation gap is formed by the levels 5S$\uparrow$ and 3AS$\downarrow$ which cross at ${\theta \approx 37^\circ}$ resulting in the conductivity approaching its  maximum. The next conductivity maximum at $\theta \approx 52^\circ$ is due to crossing of the levels 5S$\downarrow$ and 3AS$\uparrow$.
Similar curves for $\nu$=14, 10 and 6 show the same crossing angles.
{The minima at ($\nu$=9, 11, 13, and 15) are associated with transitions between different subbands, but with the same numbers of Landau levels and parallel directions of spins. The magnetic field tilt affects the amplitude of these minima very little, i.e., no level intersections are observed in these cases and we assume that anticrossings rather than crossings occur.}

Our analysis has only one fitting parameter, namely g-factor.
 The $g$-factor values determined by this procedure were $g=12$.
However, if we consider the broadening of levels in a magnetic field as~\cite{broad}
\begin{equation} \label{eqBroad}
\Delta = \left(\frac{2}{\pi} \hbar \omega_c \frac{\hbar}{\tau_q}\right)^{1/2},
%=0.13B^{1/2}, \rm{where}
\end{equation}
where $\omega_c$ is the cyclotron frequency in transverse magnetic field $B_{\perp}$, and
  $\tau_q = 4 \times 10^{-11}$~s is the quantum relaxation time, determined for this sample in Ref.~\cite{Drichko_2018}, we obtain the $g$-factor $g$=9.

Figures~\ref{fig5}~(a),~(b),~(c), and~(d) demonstrate a reasonably good agreement between theory and experiment. Some difference between experimental and theoretical angles of intersection of Landau levels in tilted magnetic fields is explained by an error in the determination of the initial sample position ($\theta=0^\circ$) {and also by not very accurate determination of the dependence of the $\sigma_1$ maxima on the tilt angle.}
The obtained $g$-factor agrees with the data available in the literature determined from transport experiments in various $n$-type GaAs/AlGaAs structures, see Ref.~\cite{g-fact} and references therein.

\section{Conclusion} \label{concl}

In the high-mobility $n$-GaAs/AlGaAs structure with a 75~nm-wide quantum well we observed a rich oscillations pattern corresponding to the integer and fractional quantum Hall effect regimes. In this work we investigated in detail  oscillations of conductivity
in magnetic fields $B_\perp = 0.5\ldots 1.5$~T. Analysis of the oscillation minima  positions in a magnetic field shows that these oscillations
%with half-integer filling factors
are caused by the two-subband electron spectrum (S and AS subbands) formed due to electrostatic repulsion of electrons. In tilted fields a strong increase of the conductivity leading to a disappearance of the minima took place at some tilt angles different for different minima. This effect is shown to be caused not by a variation of $\Delta_\text{SAS}$ with $B_{\parallel}$ which is very weak in the studied structure but by crossings of the Landau levels from the S and AS subbands. The developed theory demonstrates that the crossings in the range $B_{\parallel} =0\ldots 1.5$~T occur due to different dependencies of cyclotron energies in the subbands on the parallel field. Namely, the cyclotron energy in the S (AS) subband decreases (increases) when $B_{\parallel}$ rises. We estimated the value of the electron $g$-factor by fitting theoretical dependences to experimental data.

\acknowledgments

The authors would like to thank Yu. M. Galperin for careful reading the manuscript and providing useful comments, S. A. Tarasenko for useful discussions, and E.~Palm, T. Murphy, J.-H. Park, and G. Jones for technical assistance. Partial support from
program ``Physics and technology of nanostructures, nanoelectronics and diagnostics'' of the Presidium of RAS (I.~Y.~S.)
 and the Russian Foundation for Basic Research projects
% 16-02-0037517
19-02-00124 (I.~L.~D.) and 19-02-00095 (L.~E.~G., M.~O.~N.)
is gratefully acknowledged. M.~O.~N. and L.~E.~G. thank the Foundation for advancement of theoretical physics and mathematics "BASIS". The National High Magnetic Field Laboratory is supported by the National Science Foundation through NSF/DMR-1157490 and NSF/DMR-1644779 and the State of Florida. The work at Princeton was supported by Gordon and Betty Moore Foundation through the EPiQS initiative Grant GBMF4420, and by the NSF MRSEC Grant DMR-1420541.

\section*{References}
%\bibliography{GaAs_tilt_IOP}% Produces the bibliography via BibTeX.
%

\end{document}